\documentclass{article}
\usepackage{amsmath,amssymb,pxfonts}
\newtheorem{assu}{Condition}[section]
\newtheorem{defn}{Definition}[section]
\newtheorem{thm}{Theorem}[section]
\newtheorem{lem}{Lemma}[section]
\newtheorem{rem}{Remark}[section]
\newtheorem{prop}{Proposition}[section]

\title{Normalization for Implied Volatility}
\author{Masaaki Fukasawa\thanks{
Center for the Study of Finance and Insurance, Osaka University, 
}}
\begin{document}
\maketitle
\begin{abstract}
We study specific nonlinear transformations of
the Black-Scholes implied volatility to show
remarkable properties of the volatility surface.
Model-free bounds on the implied volatility skew are given.
Pricing formulas for  the  European options which are written 
in terms of the implied volatility are given.
In particular, we prove elegant formulas 
for the fair strikes of the variance swap and the gamma
 swap.
\end{abstract}

\section{Introduction}
This study is motivated by 
an elegant formula (11.5) of Gatheral~\cite{Gatheral}:
\begin{equation}\label{mot}
-2\mathbb{E}[\log(S_T/F)] = 
\int \sigma(g_2(z))^2 \phi(z)\mathrm{d}z,
\end{equation}
where $F$ is the forward price of the asset $S_T$,
$\sigma$ is the Black-Scholes implied volatility as a function of
log moneyness $k=\log(K/F)$ with strike $K$ and maturity $T$, 
and $g_2$ is the inverse function of the transformation $k \mapsto -
d_2(k,\sigma(k))$.
Here, we denote by $\phi$ the standard normal density and define 
$d_2$ as 
\begin{equation*}
d_2(k,\sigma) = -\frac{k}{\sigma} - \frac{\sigma}{2}.
\end{equation*}
This formula was essentially found by 
Morokoff, Akesson and Zhou~\cite{MAZ} and used in
 Chriss and Morokoff~\cite{CM}.
A rigorous treatment is given in Carr and Lee~\cite{CL}.
Not only to prove it but also to ensure that the formula
itself is well-defined, 
the preceding studies assumed, implicitly or explicitly, that the
mapping $k \mapsto -d_2(k,\sigma(k))$ is increasing.
This monotonicity is not trivial because it involves the mapping $ k \mapsto \sigma (k)$.
In this article, we show that it
is in fact increasing under the minimal no-arbitrage condition and enjoys other nice properties 
which imply in particular model-free bounds on
the implied volatility skew, that is,
the first derivative of $\sigma$.
The formula (\ref{mot}), which is known to give the fair strike of the variance swap,
is proved in an extended form.
We find also its counterpart for the fair strike of the gamma swap which
is seemingly new:
it holds that
\begin{equation}\label{gamma}
2\mathbb{E}[\log(S_T/F)S_T/F] = 
\int \sigma(g_1(z))^2 \phi(z)\mathrm{d}z.
\end{equation}
Here $g_1$ is the inverse function of the mapping
$k \mapsto -d_1(k,\sigma(k))$, which is also shown to be increasing,
where $d_1(k,\sigma) = d_2(k,\sigma) + \sigma$.
Moreover, we show that the functions $g_1$ and $g_2$ completely
characterize the distribution of $S_T$ in the following sense:
it holds for a given absolutely continuous function $\Psi$
that
\begin{equation*}
\mathbb{E}[\Psi(\log(S_T/F))]
= \int \left\{\Psi(g_2(z)) - \Psi^\prime(g_2(z)) +
\Psi^\prime(g_1(z))e^{-g_1(z)} \right\} \phi(z)\mathrm{d}z.
\end{equation*}
The results are model-independent and
directly useful in practice.
We present  basic results including the monotonicity of $d_2$ in
Section~2.
Then we introduce the normalized Black-Scholes implied volatilities in
Section~3.
Pricing formulas for the European options are given in Section~4.

\section{Basic results}
Let a nonnegative random variable $S_T$ stand for an asset price at a fixed future time $T>0$.
We assume the following condition to hold throughout this article:
\begin{assu} 
There exists a probability measure $\mathbb{E}$ such that
\begin{equation*}
P(K) = \mathbb{E}[(K-S_T)_+], \ \ F = \mathbb{E}[S_T], \ \
 \mathbb{E}[S_T=F] < 1
\end{equation*}
for all $K> 0$, where $P(K)$ is the undiscounted price of the put
 option with strike $K$ and maturity $T$ written on the asset and $F$ is the $T$-expiry
  forward price of the same asset.
\end{assu}

\begin{defn}
The (undiscounted) Black-Scholes put price is a function of
$k \in \mathbb{R}$ and $\sigma \in (0,\infty)$ 
defined as
\begin{equation*}
P_{\mathrm{BS}}(k,\sigma)
= F e^{k}\Phi(-d_2(k,\sigma))- F\Phi(-d_1(k,\sigma)),
\end{equation*}
where 
\begin{equation*}
d_2(k,\sigma) = \frac{-k -\sigma^2/2}{\sigma}, \ \ 
d_1(k,\sigma) = d_2(k,\sigma) + \sigma.
\end{equation*}
\end{defn}
\begin{defn}
The Black-Scholes implied volatility is
a function of $k \in \mathbb{R}$ defined as
\begin{equation*}
\sigma(k) = P_{\mathrm{BS}}(k,\cdot)^{-1}(P(Fe^k)),
\end{equation*}
or equivalently,
\begin{equation*}
P_{\mathrm{BS}}(k,\sigma(k)) = P(Fe^k).
\end{equation*}
\end{defn}
Recall that $P_{\mathrm{BS}}$ is an increasing function of $\sigma$  for
fixed $k \in \mathbb{R}$.
Note also that $(K-F)_+ \leq  P(K) < K$ by Jensen's inequality
and 
\begin{equation*}
\lim_{\sigma\to 0} P_{\mathrm{BS}}(k,\sigma) = F(e^k-1)_+, \ \ 
\lim_{\sigma\to \infty} P_{\mathrm{BS}}(k,\sigma) = Fe^k,
\end{equation*}
so that the Black-Scholes implied volatility is well-defined.
\begin{defn}
The first and second normalizing transformations (of log-moneyness)
are functions $f_1$ and $f_2$ on $\mathbb{R}$ defined as
$f_1(k) = -d_1(k,\sigma(k))$ and 
$f_2(k) = -d_2(k,\sigma(k))$ respectively for $k\in \mathbb{R}$;
more specifically,
\begin{equation*}
f_1(k) = \frac{k}{\sigma(k)}-\frac{\sigma(k)}{2}, \ \ 
f_2(k) = \frac{k}{\sigma(k)}+\frac{\sigma(k)}{2}.
\end{equation*}
\end{defn}
Notice that $\sigma(0) >  0$ by the condition $\mathbb{E}[S_T=F]<1$.
Therefore we can define as $f(k) = \pm \infty$ according to the
sign of $k$ if $\sigma(k)=0$. 
Denote by  $D_{\mathrm{BS}}(K)$ and $D_{\pm}(K)$ the 
functions of $K = Fe^k$ defined as
\begin{equation*}
D_{\mathrm{BS}}(K) = \frac{1}{K}\frac{\partial P_{\mathrm{BS}}}{\partial k}(k,
\sigma)|_{\sigma=\sigma(k)}
=  \Phi(f_2(k))
\end{equation*}
and
\begin{equation*}
D_+(K) = \mathbb{E}[K\geq S_T], \ \ 
D_-(K) = \mathbb{E}[K>S_T].
\end{equation*}
respectively.  We will assume sometimes, but not always, conditions below:
\begin{assu}\label{lh}
It holds that $\mathbb{E}[S_T=0] < 1/2$.
\end{assu}
\begin{assu}\label{l0}
It holds that $\mathbb{E}[S_T=0] = 0$.
\end{assu}
\begin{assu}\label{lp}
There exists $p>0$ such that $\mathbb{E}[S_T^{-p}] < \infty$.
\end{assu}
\begin{assu}\label{lpp1}
There exists $p>0$ such that $\mathbb{E}[S_T^{1+p}] < \infty$.
\end{assu}
\begin{assu}\label{dens} 
The law of $S_T$ under $\mathbb{E}$ has a density.
\end{assu}
Denote by $\partial_+^x$, $\partial_-^x$ the right and left derivative
operators with respect $x$. We omit $x$ when the operand has only one variable.
Then, it holds that
\begin{equation} \label{seven}
\begin{split}
D_{\pm}(K) &= \partial_{\pm}P(K) 
\\
& = D_{\mathrm{BS}}(K) 
+ \frac{1}{K}  
\partial_{\pm}^\sigma P_{\mathrm{BS}}
(\log(K/F),\sigma(\log(K/F)))
\partial_{\pm} \sigma (\log(K/F))
\\&=
D_{\mathrm{BS}}(K) + \phi(-d_2(\log(K/F),\sigma(\log(K/F))))
\partial_{\pm} \sigma (\log(K/F)).
\end{split}
\end{equation}

\begin{lem}\label{d2lem}
It holds for all $k \in \mathbb{R}$ that
\begin{equation*}
f_2(k)\partial_{\pm }\sigma (k) < 1.
\end{equation*}
\end{lem}
{\it Proof: }
The inequality is trivial when $f_2(k)=0$.
If $f_2(k) > 0$, it follows from (\ref{seven}) that
\begin{equation*}
f_2(k)\partial_{\pm}\sigma(k) =
f_2(k)\frac{D_{\pm}(Fe^k)-D_{\mathrm{BS}}(Fe^k)}{\phi(f_2(k))}
\leq
f_2(k)\frac{1 - \Phi(f_2(k))}{\phi(f_2(k))} < 1.
\end{equation*}
Here we used the fact that $0\leq D_{\pm}(K) \leq 1$ by definition 
and a well-known estimate
\begin{equation}\label{ntail}
1-\Phi(x) <  x^{-1}\phi(x), \ \ x > 0.
\end{equation}
For the case $f_2(k) < 0$, we have
\begin{equation*}
\begin{split}
&f_2(k)\partial_{\pm}\sigma(k) =
f_2(k)\frac{D_{\pm}(Fe^k)-D_{\mathrm{BS}}(Fe^k)}{\phi(f_2(k))}
\\\leq &
- f_2(k)\frac{\Phi(f_2(k))}{\phi(f_2(k))} =
- f_2(k)\frac{1-\Phi(-f_2(k))}{\phi(-f_2(k))} < 1.
\end{split}
\end{equation*}
\hfill////

\begin{prop} \label{inclem1}
The first normalizing transformation $f_1$ is an increasing function.
\end{prop}
{\it Proof: }
This follows from Lemma~\ref{d2lem} because
\begin{equation*}
\partial_{\pm }f_1 (k)=
\frac{1}{\sigma(k)}\left\{
1 - \partial_{\pm}\sigma(k) f_2(k)
\right\}
\end{equation*}
by a simple calculation.
\hfill////

\begin{lem}\label{d1lem}
It holds for all $k$ with  $f_1(k) \leq  0$ that
\begin{equation*}
f_1(k) \partial_{\pm}\sigma(k)  < 1.
\end{equation*}
\end{lem}
{\it Proof: }
By definition, it holds that for all $K > 0$,
\begin{equation*}
KD_{\pm}(K) \geq P(K).
\end{equation*}
Combining this and (\ref{seven}), we have
\begin{equation*}
F\Phi(-d_1(k,\sigma(k))) +
 K\phi(-d_2(k,\sigma(k)))
\partial_{\pm}\sigma(k) \geq 0
\end{equation*}
with $k=\log(K/F)$. Since $K\phi(-d_2) = F\phi(-d_1)$,
we obtain from (\ref{ntail}),
\begin{equation*}
\partial_{\pm}\sigma(k) \geq
-\frac{1-\Phi(d_1(k,\sigma(k)))}{\phi(d_1(k,\sigma(k)))}
> -\frac{1}{d_1(k,\sigma(k))}.
\end{equation*}
\hfill////

\begin{prop}\label{inclem}
The second normalizing transformation $f_2$ 
is an increasing function.
\end{prop}
{\it Proof: }
By definition,
\begin{equation} \label{fp}
\partial_{\pm}f_2 (k) = \frac{1}{\sigma(k)}\left\{
1 -\partial_{\pm}\sigma(k)\frac{k}{\sigma(k)}
\right\} + \frac{1}{2}\partial_{\pm}\sigma(k)
= \frac{1}{\sigma(k)}\left\{
1 -\partial_{\pm}\sigma(k)f_2(k) \right\} + 
\partial_{\pm}\sigma(k).
\end{equation}
Hence, by Lemma~\ref{d2lem}, we have
\begin{equation*}
\partial_{\pm}f_2 (k)
> \partial_{\pm}\sigma(k).
\end{equation*}
It suffices then to treat the case $\partial_{\pm} \sigma (k) < 0$.
By rewriting (\ref{fp}), we have
\begin{equation} \label{f2f1}
\partial_{\pm}f_2 (k) = 
\frac{1}{\sigma(k)}\left\{
1 + \partial_{\pm}\sigma(k) d_1(k,\sigma(k))
\right\}.
\end{equation}
If $d_1(k,\sigma(k)) < 0$, we have $\partial_{\pm}f_2 (k)  > 0$
under $\partial_{\pm} \sigma (k) < 0$.
If $d_1(k,\sigma(k)) \geq 0$, we can use Lemma~\ref{d1lem}
to obtain the same inequality. 
\hfill////

\begin{lem} \label{d1lem2}
It holds for all $k \in \mathbb{R}$ that 
\begin{equation*}
f_1(k) \partial_{\pm}\sigma(k)  < 1.
\end{equation*}
\end{lem}
{\it Proof: } This follows from (\ref{f2f1}) and
Proposition~\ref{inclem}. \hfill////

\begin{lem}\label{stail}
It holds that $f_1(k) \leq -\sqrt{2|k|}$ for all $k\leq 0$.
\end{lem}
{\it Proof: }
This is because the arithmetic mean exceeds the geometric mean.
\hfill////

\begin{prop}\label{skews}
The mapping $k \to \sqrt{2|k|} -\sigma(k)$ is decreasing on
 $(-\infty,0]$; 
it holds for all $k < 0$ that
\begin{equation*}
\partial_{\pm}\sigma(k) > - \frac{1}{\sqrt{2|k|}}.
\end{equation*}
\end{prop}
{\it Proof: } This follows from
Lemmas~\ref{d1lem} and \ref{stail}.
\hfill////

\begin{prop}\label{equivl}
It holds that
\begin{equation*}
\mathbb{E}[S_T=0] = \Phi(\lim_{k\to -\infty} f_2(k)).
\end{equation*}
In particular, Condition~\ref{lh} holds if and only if there exists $k^\ast < 0$ such
 that $f_2(k^\ast) <  0$.
This condition is also equivalent to that there exists $k^\ast < 0$ such
 that
\begin{equation*}
\sigma(k) < \sigma(k^\ast) +  \sqrt{2|k|}-\sqrt{2|k^\ast|} < \sqrt{2|k|}
\end{equation*}
 for all  $k < k^\ast$.
\end{prop}
{\it Proof: } By definition,
\begin{equation*}
\mathbb{E}[(1-S_T/K)_+] = \Phi(f_2(k)) - F \Phi(f_1(k))/K
\end{equation*}
and the left hand side converges to $\mathbb{E}[S_T=0]$ as $K\to 0$.
The first identity follows from
\begin{equation*}
F \Phi(f_1(k))/K = e^k
 \phi(-f_1(k))\frac{1-\Phi(-f_1(k))}{\phi(-f_1(k))} < 
\frac{1}{2\sqrt{\pi |k|}} \to 0 \text{ as } k \to -\infty
\end{equation*}
Here we used (\ref{ntail}) and Lemma~\ref{stail}.
To show the other equivalence, notice that
\begin{equation*}
\Phi(f_2(k)) < 1/2 \Leftrightarrow 
f_2(k) < 0 \Leftrightarrow \sigma(k) < \sqrt{2|k|}
\end{equation*}
for $k <  0$ and use Proposition~\ref{skews}.
\hfill////

\begin{prop}
Condition~\ref{l0} holds if and only if $\lim_{k \to -\infty}f_2(k) =
 -\infty$.
If one of these equivalent conditions holds, then
\begin{equation*}
\sqrt{2|k|}-\sigma(k) \to \infty \text{ as } k \to -\infty.
\end{equation*}
\end{prop}
{\it Proof: }
The first assertion follows immediately from Proposition~\ref{equivl}.
Suppose that the decreasing function $a(k):=\sqrt{2|k|}-\sigma(k)$ is
bounded above. Then
\begin{equation*}
2f_2(k) = \sqrt{2|k|}\left\{
1- \frac{a(k)}{\sqrt{2|k|}} - \left(1- \frac{a(k)}{\sqrt{2|k|}}\right)^{-1} 
\right\}
= \sqrt{2|k|}\left\{
-\frac{2a(k)}{\sqrt{2|k|}} + O(\frac{1}{2|k|})
\right\}
\end{equation*}
as $k\to -\infty$, which implies $f_2$ is bounded below. \hfill////

\begin{rem} \label{rem1} \upshape
Proposition~\ref{equivl} is a slight refinement of Lemma~3.3 of Lee~\cite{Lee}, where he showed Condition~\ref{lh} holds if and only if  there
 exists $k^\ast < 0$ such that $\sigma(k) < \sqrt{2|k|}$ for all $k < k^\ast$.
\end{rem}

\begin{rem}\upshape
Putting $V(k)= \sigma(k)^2$, 
Theorem~5.1 of Rogers and Tehranchi~\cite{RT} gives a lower bound
\begin{equation*}
\partial_-V(k) > - 4
\end{equation*}
for $k \leq 0$.
By Proposition~\ref{skews}, it holds for all $k, k_0$ with $k \leq k_0 \leq 0$ that 
\begin{equation} \label{Vpb}
\partial_{\pm}V(k) > -2 \frac{\sigma(k)}{\sqrt{2|k|}} > 
 -2 \left\{
1 - \frac{\sqrt{2|k_0|} -\sigma(k_0) }{\sqrt{2|k|}}
\right\}.
\end{equation}
If $k_0 < k^\ast < 0$, then
$\sigma(k_0) < \sqrt{2|k_0|}$, so that
(\ref{Vpb}) gives a sharper bound for $k < k^\ast$ under Condition~\ref{lh}.
\end{rem}

\begin{lem}\label{ltail}
It holds that $f_2(k) \geq \sqrt{2k}$ for all $k\geq 0$.
\end{lem}
{\it Proof: }
This is because the arithmetic mean exceeds the geometric mean.
\hfill////

\begin{prop}\label{skewl}
The mapping $\sqrt{2k}-\sigma(k)$ is increasing on $[0,\infty)$;
it holds for all $k>0$ that,
\begin{equation*}
\partial_{\pm}\sigma(k) <  \frac{1}{\sqrt{2k}}.
\end{equation*}
\end{prop}
{\it Proof: } This follows from
Lemmas~\ref{d2lem} and \ref{ltail}.
\hfill////

\begin{prop}\label{equivr}
It holds that
\begin{equation*}
 f_1(k)  \to  \infty, \ \ \sqrt{2k} -\sigma(k) \to \infty \text{ as } k
  \to \infty
\end{equation*}
and there exists $k^\ast > 0$ such that
\begin{equation*}
\sigma(k) < \sigma(k^\ast) + \sqrt{2k}-\sqrt{2k^\ast} < \sqrt{2k}
\end{equation*}
for all $k> k^\ast$.
\end{prop}
{\it Proof: }
The first claim on the divergences was given in 
Theorem~5.5 of Rogers and Tehranchi~\cite{RT} but here we give its proof
for the readers' convenience.
By definition, or ``Call-Put Parity'',
\begin{equation*}
\mathbb{E}[(S_T-K)_+] = 
\mathbb{E}[(K-S_T)_+] + F-K = F\Phi(-f_1(k)) -K\Phi(-f_2(k)).
\end{equation*}
The left hand side goes to $0$ as $K \to \infty$ and
\begin{equation*}
K\Phi(-f_2(k)) = Fe^k \phi(f_2(k)) \frac{1-\Phi(f_2(k))}{\phi(f_2(k))}
 < \frac{F}{2\sqrt{\pi k}} \to 0 \text{ as } k \to \infty.
\end{equation*}
Here we used (\ref{ntail}) and Lemma~\ref{ltail}.
Hence we have the first divergence. 
The second divergence follows from the first. In fact 
if the increasing function  $a(k):=\sqrt{2k}-\sigma(k)$ is
 bounded above,
then
\begin{equation*}
2f_1(k) = \sqrt{2k}\left\{ \left(1-\frac{a(k)}{\sqrt{2k}}\right)^{-1} -
\left(1-\frac{a(k)}{\sqrt{2k}}\right)\right\}
= \sqrt{2k}\left(\frac{2a(k)}{\sqrt{2k}} + O(\frac{1}{2k})\right)
\end{equation*}
as $k\to \infty$, which contradicts the divergence of $f_1$.
To see the last inequality, notice
that
\begin{equation*}
f_1(k)>0 \Leftrightarrow \sigma(k) < \sqrt{2k}
\end{equation*}
for $k>0$ and use Proposition~\ref{skewl}.
\hfill////

\begin{rem} \label{rem2} \upshape
Proposition~\ref{equivr} gives a slight refinement of
Lemma~3.1 of Lee~\cite{Lee}, where he showed that there exists $k^\ast > 0$
 such that $\sigma(k) < \sqrt{2k}$ for all  $k > k^\ast$.
Putting $V(k) = \sigma(k)^2$,
Theorem~5.1 of Rogers and Tehranchi~\cite{RT} gives a upper bound
\begin{equation*}
\partial_{+}V(k)  <  4
\end{equation*}
for $k \geq 0$. 
Theorem~5.5 of the same paper gives that
there exists $k^\ast > 0$ such that
 \begin{equation*}
\partial_{+}V(k) <  2
\end{equation*}
for all $k \geq k^\ast$.
From Proposition~\ref{skewl}, it follows for all $k, k_0$ with $k > k_0 > k^\ast > 0$ that 
\begin{equation*}
\partial_{\pm}V(k)
 <  2 \frac{\sigma(k)}{\sqrt{2k}} < 
 2 \left\{
1 - \frac{\sqrt{2k_0}-\sigma(k_0)}{\sqrt{2k}}
\right\} < 2.
\end{equation*}
Proposition~\ref{skewl} therefore gives more precise estimate.
\end{rem}

\begin{lem}
Under Condition~\ref{lp},
there exists $q \in (0,2)$ such that 
it holds for all $k<-1/(2-q)$  that
\begin{equation}\label{blp}
\sigma(k) < \sqrt{q|k|},\ \ 
f_1(k) < - \left\{
\frac{1}{\sqrt{q}} + \frac{\sqrt{q}}{2} \right\} \sqrt{|k|}, \ \ 
f_2(k) < -(\sqrt{2}-\sqrt{q})\sqrt{|k|}
\end{equation}
and 
\begin{equation*}
\partial_{\pm} \sigma (k) > - \left\{
\frac{1}{\sqrt{q}} + \frac{\sqrt{q}}{2} \right\}^{-1} \frac{1}{\sqrt{|k|}}.
\end{equation*}
\end{lem}
{\it Proof: }
The first bound on $\sigma$ was given by Lee~\cite{Lee}.
The bounds on $f_1$ and $f_2$ follow from the first by noting that
\begin{equation*}
\begin{split}
&f_1(k) = - \left\{\frac{|k|}{\sigma(k)} +
	    \frac{a \sigma(k)}{2}\right\}
+ \frac{(a-1)\sigma(k)}{2} < 
-\sqrt{2a|k|} + \frac{(a-1)\sqrt{q|k|}}{2},\\
&f_2(k) = f_1(k)+\sigma(k) < -\sqrt{2|k|} + \sqrt{q|k|}
\end{split}
\end{equation*}
with $a=2/q$. Here we used the fact that the arithmetic mean exceeds the
geometric mean. The last bound then follows from Lemma~\ref{d1lem}.
\hfill////

\begin{lem}
Under Condition~\ref{lpp1},
there exists $q \in (0,2)$ such that 
it holds for all $k > 1/(2-q)$  that
\begin{equation}\label{blpp1}
\sigma(k) < \sqrt{qk},\ \ 
f_1(k) > (\sqrt{2}-\sqrt{q})\sqrt{k}, \ \ 
f_2(k) >  \left\{
\frac{1}{\sqrt{q}} + \frac{\sqrt{q}}{2} \right\} \sqrt{k}
\end{equation} 
and
\begin{equation*}
\partial_{\pm} \sigma (k) <  \left\{
\frac{1}{\sqrt{q}} + \frac{\sqrt{q}}{2} \right\}^{-1} \frac{1}{\sqrt{k}}.
\end{equation*}
\end{lem}
{\it Proof: } The proof  is similar to the previous one.  \hfill////

\begin{rem} \upshape
The bounds we obtained so far are based on (\ref{ntail}). A sharper
 bound is however known.
Formula 7.1.13 from Abramowitz and Stegun~\cite{AS} implies that
\begin{equation*}
\frac{1-\Phi(x)}{\phi(x)} \leq \frac{2}{x+\sqrt{x^2 + 8/\pi}}
\end{equation*}
for $x \geq 0$. Lemma~\ref{d2lem} is then improved as
\begin{equation*}
f_2(k)\partial_{\pm}\sigma(k) \leq
 \frac{2|f_2(k)|}{|f_2(k)|+\sqrt{|f_2(k)|^2+8/\pi}}.
\end{equation*}
Lemma~\ref{d1lem} also has an improvement.
In particular in light of Lemmas~\ref{stail} and \ref{ltail}, we have for all $k \in \mathbb{R}$ that
\begin{equation*}
\mathrm{sgn}(k)\partial_{\pm}\sigma(k) \leq \frac{2}{\sqrt{2|k|} + \sqrt{2|k|+8/\pi}} <
 \sqrt{\frac{\pi}{2}} \wedge \frac{1}{\sqrt{2|k|}}.
\end{equation*}
\end{rem}

\section{Normalized Black-Scholes implied volatilities}
Here we introduce the normalized Black-Scholes implied volatilities and
give model-free bounds on them. 
They play an important role in the next section.

\begin{defn}
The first and second normalized Black-Scholes implied volatilities are functions
 on $\mathbb{R}$ defined as  $\sigma_1(z) = \sigma(g_1(z))$ and 
$\sigma_2(z) = \sigma(g_2(z))$ respectively for $z \in \mathbb{R}$,
 where
$g_1$, $g_2$ are the inverse functions of the increasing functions
 $f_1$, $f_2$ respectively.
\end{defn}

\begin{prop}
The mappings $z \mapsto z+\sigma_1(z)$, $z \mapsto z-\sigma_2(z)$ are increasing;
it holds for all $z \in \mathbb{R}$ that
\begin{equation*}
\partial_{\pm} \sigma_1 (z)  > -1, \ \ 
\partial_{\pm} \sigma_2 (z) < 1.
\end{equation*}
\end{prop}
{\it Proof: }
The second inequality follows from (\ref{fp}) and Lemma~\ref{d2lem}.
The first one follows from
\begin{equation*}
\partial_{\pm}f_1 (k)
= \frac{1}{\sigma(k)}\left\{
1 -\partial_{\pm}\sigma(k)f_1(k) \right\} -
\partial_{\pm}\sigma(k)
\end{equation*}
and Lemma~\ref{d1lem2}.
\hfill////

\begin{prop}\label{sigma1}
Put 
\begin{equation*}
\alpha_1^{\pm}(z;z_0) = -z \pm  \sqrt{\sigma_1(z_0)^2 + 2z_0 \sigma_1(z_0) +
 z^2}.
\end{equation*}
Then, it holds for $z > z_0 \geq 0$ that
\begin{equation*}
\sigma_1(z) > \alpha_1^+(z,z_0) > \sigma_1(z_0) +z_0-z.
\end{equation*}
It also holds for $z \in [0,z_0)$ that
\begin{equation*}
\sigma_1(z) < \alpha_1^+(z,z_0) < \sigma_1(z_0) +z_0-z.
\end{equation*}
Moreover, it holds for $z < z_0 \leq 0$ that
\begin{equation*}
\sigma_1(z) > \alpha_1^-(z,z_0).
\end{equation*}
Under Condition~\ref{lh}, there exists
 $z^\ast_1 < 0$ such that 
$\sigma_1(z^\ast_1) = - z^\ast_1$, \ $z^\ast_1 = f_1(g_2(0))$ and 
$\alpha_1^-(z,z_0) > 0$ for all $z \leq z_0 \leq  z^\ast_1$. 
\end{prop}
{\it Proof: }
The bounds follow from the fact that the mapping
\begin{equation*}
k \mapsto f_2(k)^2- f_1(k)^2 = 2k
\end{equation*}
is an increasing function.
For the existence of $z^\ast_1$, notice that
$\sigma_1(z^\ast_1) = - z^\ast_1$ is equivalent to
\begin{equation*}
\sigma(k_1^\ast) = \sqrt{2 |k^\ast_1|}, \ \  f_1(k^\ast_1) = z^\ast_1.
\end{equation*}
See Proposition~\ref{equivl} for the existence of such $k^\ast_1 < 0$.
 \hfill////

\begin{rem}\upshape
Because the inverse function of $f_1$, 
\begin{equation*}
g_1(z) = z\sigma_1(z) + \frac{1}{2}\sigma_1(z)^2
\end{equation*}
is an increasing function, we obtain by considering the first derivative
 of $g_1$ that 
\begin{equation*}
\partial_{\pm}\sigma_1(z) > -
 \frac{\sigma_1(z)}{\sigma_1(z)+z} 
\end{equation*}
for $z\geq z^\ast_1$ and
\begin{equation*}
\partial_{\pm}\sigma_1(z) < -
 \frac{\sigma_1(z)}{\sigma_1(z)+z} 
\end{equation*}
for $z <  z^\ast_1$. Since $\alpha^{\pm}_1(\cdot,z_0)$ satisfy the
 ordinary differential equation
\begin{equation*}
\psi^\prime(z) = - \frac{\psi(z)}{\psi(z)+z}, \ \ \psi(z) = \sigma_1(z_0)
\end{equation*}
for $\psi$, this bounds on $\partial_{\pm} \sigma_1$ result in the same bounds for $\sigma_1$
 as in Proposition~\ref{sigma1}.

\end{rem}

\begin{prop}\label{sigma2}
Put
\begin{equation*}
\alpha_2^{\pm}(z;z_0) = z \pm  \sqrt{\sigma_2(z_0)^2 - 2z_0 \sigma_2(z_0) +
 z^2}.
\end{equation*}
Then, it holds for $z < z_0 \leq 0$ that
\begin{equation*}
\sigma_2(z) > \alpha_2^+(z,z_0) > \sigma_2(z_0) - z_0 + z.
\end{equation*}
It also holds for $z \in (z_0,0]$ that
\begin{equation*}
\sigma_2(z) < \alpha_2^+(z,z_0) < \sigma_2(z_0) -z_0 + z.
\end{equation*}
Moreover, it holds for $z >  z_0 \geq 0$ that
\begin{equation*}
\sigma_2(z) > \alpha_2^-(z,z_0).
\end{equation*}
There exists
 $z^\ast_2 > 0$ such that 
$\sigma_2(z^\ast_2) =  z^\ast_2$, \ $z^\ast_2 = f_2(g_1(0))$ and 
$\alpha_2^-(z,z_0) > 0$ for all $z \geq z_0 \geq  z^\ast_2$.
\end{prop}
{\it Proof: }
This also follows from the fact that the mapping
\begin{equation*}
k \mapsto f_2(k)^2- f_1(k)^2 = 2k
\end{equation*}
is an increasing function. 
For the existence of $z^\ast_2$, notice that
$\sigma_2(z^\ast_2) =  z^\ast_2$ is equivalent to
\begin{equation*}
\sigma(k_2^\ast) = \sqrt{2 k^\ast_2}, \ \  f_2(k^\ast_2) = z^\ast_2.
\end{equation*}
See Proposition~\ref{equivr} for the existence of such $k^\ast_2 > 0$.
\hfill////

\begin{rem}\upshape
Because the inverse function of $f_2$, 
\begin{equation*}
g_2(z) = z\sigma_2(z) - \frac{1}{2}\sigma_2(z)^2
\end{equation*}
is an increasing function, we obtain by considering the first derivative
 of $g_2$ that 
\begin{equation*}
\partial_{\pm}\sigma_2(z) > 
 \frac{\sigma_2(z)}{\sigma_2(z)-z} 
\end{equation*}
for $z\geq z^\ast_2$ and
\begin{equation*}
\partial_{\pm}\sigma_2(z) < 
 \frac{\sigma_2(z)}{\sigma_2(z)-z} 
\end{equation*}
for $z <  z^\ast_2$. Since $\alpha^{\pm}_2(\cdot,z_0)$ satisfy the
 ordinary differential equation
\begin{equation*}
\psi^\prime(z) = \frac{\psi(z)}{\psi(z)-z}, \ \ \psi(z) = \sigma_2(z_0)
\end{equation*}
for $\psi$, this bounds on $\partial_{\pm} \sigma_2$ result in the same bounds for $\sigma_2$
 as in Proposition~\ref{sigma2}.
\end{rem}

\begin{rem} \upshape
Here we give bounds on $g_1$ and $g_2$ under additional conditions. 
By (\ref{blp}), Condition~\ref{lp} implies that
there exists $q \in (0,2)$ such that it holds for all $z < -1/(2-q)$ that
\begin{equation}\label{blp2}
g_1(z) >  - \left\{ \frac{1}{\sqrt{q}} +
	     \frac{\sqrt{q}}{2}\right\}^{-1}z^2, \ \ 
g_2(z) > - (\sqrt{2}-\sqrt{q})^{-1}z^2.
\end{equation}
By (\ref{blpp1}), Condition~\ref{lpp1} implies that
there exists $q \in (0,2)$ such that it holds for all $z > 1/(2-q)$ that
\begin{equation}\label{blpp12}
g_1(z) <  (\sqrt{2}-\sqrt{q})^{-1}z^2, \ \ 
g_2(z) < \left\{ \frac{1}{\sqrt{q}} +
	     \frac{\sqrt{q}}{2}\right\}^{-1}z^2.
\end{equation}
\end{rem}

\section{Pricing formulas for the European options}
Here we present pricing formulas for the European options which 
extend (\ref{mot}).

\begin{lem} \label{nultail}
Let $\Psi$ be a function of polynomial growth.
It holds that
\begin{equation*}
\begin{split}
&\lim_{k \to  \infty} |\Psi(k)|
\sigma(k) \phi(f_2(k))
= 0, \ \ 
\lim_{k\to  \infty} |\Psi(k)|
\left|\partial_{\pm}\sigma (k)\right|
\phi(f_2(k)) = 0,\\
&\lim_{k \to  -\infty} |\Psi(k)|
\sigma(k) \phi(f_1(k))
= 0, \ \ 
\lim_{k\to  - \infty} |\Psi(k)|
\left|\partial_{\pm}\sigma (k)\right|
\phi(f_1(k)) = 0.
\end{split}
\end{equation*}
Under Condition~\ref{lp},
it holds that 
\begin{equation*}
\lim_{k \to - \infty} |\Psi(k)|
\sigma(k) \phi(f_2(k))
= 0, \ \ 
\lim_{k\to- \infty} |\Psi(k)|
\left|\partial_{\pm}\sigma (k)\right|
\phi(f_2(k)) = 0.
\end{equation*}
Under Condition~\ref{lpp1},
it holds that 
\begin{equation*} 
\lim_{k \to  \infty} |\Psi(k)|
\sigma(k) \phi(f_1(k))
= 0, \ \ 
\lim_{k\to  \infty} |\Psi(k)|
\left|\partial_{\pm}\sigma (k)\right|
\phi(f_1(k)) = 0.
\end{equation*}
\end{lem}
{\it Proof: }
Use
Propositions~\ref{equivl} and \ref{equivr}, 
Lemmas~\ref{stail} and \ref{ltail}, (\ref{blp}) and (\ref{blpp1}).
\hfill////

Note that under Condition~\ref{dens},
$\sigma$ is continuously
differentiable and its derivative is absolutely continuous.
We can put therefore  $D= D_+ = D_-$, $D^\prime= \partial_{+}D = \partial_- D$ and 
$\sigma^\prime = \partial_+\sigma =  \partial_-\sigma$.

\begin{thm}\label{main}
Let $\Psi$ be a differentiable function
such that the derivative  $\Psi^{\prime}$ is absolutely continuous and of polynomial growth.
Under Conditions~\ref{lp} and \ref{dens}, it holds that
\begin{equation*}
\begin{split}
\mathbb{E}[\Psi(\log(S_T/F))] 
= & 
\int_{-\infty}^{\infty} \left\{
\Psi(g_2(z)) - \Psi^\prime(g_2(z))\left\{
g_2(z) + \frac{1}{2}\sigma_2(z)^2
\right\}
\right\}\phi(z)\mathrm{d}z \\
& + \int_{-\infty}^{\infty} \Psi^{\prime \prime}(k)\sigma(k)\phi(f_2(k))\mathrm{d}k.
\end{split}
\end{equation*}
\end{thm}
{\it Proof: } 
Since the density of $S_T$ is given by $D^\prime$, 
\begin{equation*}
\mathbb{E}[\Psi(\log(S_T/F))] = \int_0^{\infty} 
\Psi(\log(K/F)) D^\prime(K)
\mathrm{d}K = 
\int_{-\infty}^{\infty} \Psi(k) D^\prime
(Fe^k)
Fe^k\mathrm{d}k.
\end{equation*}
Using (\ref{seven}), we have
\begin{equation*}
D^\prime(Fe^k) =
\frac{1}{Fe^k}
\phi(f_2(k)) \left\{ \frac{\mathrm{d} f_2}{\mathrm{d} k}(k)
\left( 1 -f_2(k) \sigma^\prime(k)
\right) + \frac{\mathrm{d}^2 \sigma}{\mathrm{d} k^2}(k)\right\}
\end{equation*}
Since
\begin{equation*}
\frac{\mathrm{d}}{\mathrm{d} k} \phi(f_2(k))
= -\phi(f_2(k))f_2(k)\frac{\mathrm{d}f_2}{\mathrm{d}k}(k),
\end{equation*}
we have
\begin{equation*}
\begin{split}
-\int_{-\infty}^{\infty} 
\Psi(k)\phi(f_2(k))f_2(k)&\frac{\mathrm{d}f_2}{\mathrm{d}k}(k)
\sigma^\prime(k)
\mathrm{d}k
= \left[ \Psi(k) \sigma^\prime(k)
\phi(f_2(k))
\right]_{-\infty}^{\infty} \\
&- \int_{-\infty}^{\infty}
\left\{\Psi^\prime(k)\sigma^\prime(k)
+ \Psi(k)\frac{\mathrm{d}^2\sigma}{\mathrm{d}k^2}(k) \right\}
\phi(f_2(k))\mathrm{d}k.
\end{split}
\end{equation*}
Hence, by Lemma~\ref{nultail},
\begin{equation*}
 \mathbb{E}[\Psi(\log(S_T/F))]
= \int_{-\infty}^{\infty}
\phi(f_2(k))\left\{
\Psi(k) \frac{\mathrm{d} f_2}{\mathrm{d} k}(k)
-\Psi^\prime(k) \sigma^\prime(k)
\right\} \mathrm{d}k.
\end{equation*}
Since
\begin{equation*}
\begin{split}
&\int_{-\infty}^{\infty}
\phi(f_2(k)) \Psi^\prime(k)\sigma^\prime(k)
\mathrm{d}k\\
&= \left[
\phi(f_2(k))\Psi^\prime(k)\sigma(k)
\right]_{-\infty}^{\infty} - \int_{-\infty}^{\infty}
\frac{\mathrm{d}(\phi\Psi^\prime \circ g_2)}{\mathrm{d}z}(f_2(k)) 
\frac{\mathrm{d} f_2}{\mathrm{d} k}(k)
\sigma(k)
\mathrm{d}k
\\
&= -\int_{-\infty}^{\infty}
\left\{
\Psi^{\prime \prime}(k)g^\prime(f_2(k)) \phi(f_2(k))
- \Psi^\prime(k)f_2(k)\phi(f_2(k))
\right\}
\frac{\mathrm{d} f_2}{\mathrm{d} k}(k)
\sigma(k)
\mathrm{d}k,
\end{split}
\end{equation*}
we obtain
\begin{equation*}
\begin{split}
&\mathbb{E}[\Psi(\log(S_T/F))] \\
&= \int_{-\infty}^{\infty}
\left\{
\Psi(g_2(z)) - \Psi^\prime(g_2(z))z\sigma(g_2(z)) + \Psi^{\prime \prime}
(g_2(z)) \sigma(g_2(z))g_2^\prime(z) 
\right\}
\phi(z) \mathrm{d}z\\
&= \int_{-\infty}^{\infty}
\left\{
\Psi(g_2(z)) - \Psi^\prime(g_2(z))z\sigma_2(z) 
\right\}
\phi(z) \mathrm{d}z + 
\int_{-\infty}^{\infty} \Psi^{\prime \prime}
(k) \sigma(k)\phi(f_2(k))\mathrm{d}k.
\end{split}
\end{equation*}
By definition,
\begin{equation*}
g_2(z) - z\sigma_2(z) = 
k -f_2(k)\sigma(k) = 
k + d_2(k,\sigma(k))\sigma(k) = - \frac{\sigma(k)^2}{2},
\end{equation*}
which completes the proof.
\hfill////
\begin{thm}\label{mainp}
Let $\Psi$ be a differentiable function
such that the derivative  $\Psi^{\prime}$ is absolutely continuous and of polynomial growth.
Under Conditions~\ref{lpp1} and \ref{dens},
it holds that
\begin{equation*}
\begin{split}
\mathbb{E}[\Psi(\log(S_T/F))S_T/F] 
= & 
\int_{-\infty}^{\infty} \left\{
\Psi(g_1(z)) - \Psi^\prime(g_1(z))\left\{
g_1(z) - \frac{1}{2}\sigma_1(z)^2
\right\}
\right\}\phi(z)\mathrm{d}z \\
& + \int_{-\infty}^{\infty} \Psi^{\prime \prime}(k)\sigma(k)\phi(f_1(k))\mathrm{d}k.
\end{split}
\end{equation*}
\end{thm}
{\it Proof: }
Repeat the same argument as in the proof of Theorem~\ref{main} by
replacing  $\Psi$ with $k \mapsto e^k\Psi(k)$. 
Here we use $e^k\phi(f_2(k)) = \phi(f_1(k))$ and
\begin{equation*}
\frac{\mathrm{d}f_2}{\mathrm{d}k}(k) = 
\frac{\mathrm{d}f_1}{\mathrm{d}k}(k) + \sigma^\prime(k)
\end{equation*}
to obtain
\begin{equation*}
 \mathbb{E}[\Psi(\log(S_T/F))S_T/F]
= \int_{-\infty}^{\infty}
\phi(f_1(k))\left\{
\Psi(k) \frac{\mathrm{d} f_1}{\mathrm{d} k}(k)
-\Psi^\prime(k) \sigma^\prime(k)
\right\} \mathrm{d}k.
\end{equation*}
\hfill////

\begin{rem} \upshape
Letting $\Psi(k) = k$ in Theorems~\ref{main} and \ref{mainp}, 
we obtain (\ref{mot}) and (\ref{gamma}) respectively.
%
These values coincide with the fair strikes of the variance swap and the gamma
 swap respectively in continuous semimartingale setting.
To see this, let $S$ be a continuous semimartingale and $S^0$ be a 
deterministic process of locally bounded variation, which stand for a risky asset
 price process and a risk-free asset price process respectively. 
Put $\tilde{S}_t = S_t/S_0$, $\tilde{S}^0_t = S^0_t/S^0_0$ and
$S^\ast_t = \tilde{S}_t/\tilde{S}^0_t$.
We may suppose that $S^\ast$ is a local martingale
under the risk neutral measure $\mathbb{E}$. Then, 
$S_T/F=S_T/\mathbb{E}[S_T] = S^\ast_T$. By It$\hat{\text{o}}$'s formula,
 we have
\begin{equation*}
-2\mathbb{E}[\log(S^\ast_T)] = \mathbb{E}\left[\int_0^T
\frac{\mathrm{d}\langle S^\ast\rangle_t}{|S_t^\ast|^2}
\right] 
=  \mathbb{E}\left[\int_0^T
\frac{\mathrm{d}\langle S \rangle_t}{|S_t|^2}
\right]  = \mathbb{E}[\langle \log(S) \rangle_T]
\end{equation*}
and
\begin{equation*}
2\mathbb{E}[\log(S^\ast_T)S^\ast_T] = \mathbb{E}\left[\int_0^T
\frac{\mathrm{d}\langle S^\ast\rangle_t}{S_t^\ast}
\right] 
=  \mathbb{E}\left[\int_0^T
S^\ast_t \mathrm{d}\langle \log(S) \rangle_t
\right].
\end{equation*}
\end{rem}

\begin{thm}
Let $\Psi$ be an absolutely continuous function with derivative  $\Psi^{\prime}$ of polynomial growth.
Under Conditions~\ref{lp} and \ref{dens},
it holds that
\begin{equation*}
\mathbb{E}[\Psi(\log(S_T/F))] = 
\int_{-\infty}^{\infty} \left\{
\Psi(g_2(z)) - \Psi^\prime(g_2(z)) + \Psi^\prime(g_1(z))e^{-g_1(z)}
     \right\} \phi(z)\mathrm{d}z.
\end{equation*}
\end{thm}
{\it Proof: }
We have
\begin{equation*}
 \mathbb{E}[\Psi(\log(S_T/F))]
= \int_{-\infty}^{\infty}
\phi(f_2(k))\left\{
\Psi(k) \frac{\mathrm{d} f_2}{\mathrm{d} k}(k)
-\Psi^\prime(k) \sigma^\prime(k)
\right\} \mathrm{d}k
\end{equation*}
by the same argument as in Theorem~\ref{main}.
Now, notice that
\begin{equation*}
\begin{split}
\int \Psi^\prime(g_1(z))e^{-g_1(z)}\phi(z)\mathrm{d}z
=& \int
\Psi^\prime(k)e^{-k}\phi(f_1(k))\frac{\mathrm{d}f_1}{\mathrm{d}k}(k)
 \mathrm{d}k\\
=& \int
\Psi^\prime(k)\phi(f_2(k))\left\{\frac{\mathrm{d}f_2}{\mathrm{d}k}(k) -
 \sigma^\prime(k) \right\}
 \mathrm{d}k\\
=& \int \Psi^\prime(g_2(z))\phi(z) \mathrm{d}z - \int \phi(f_2(k)) \Psi^\prime(k)\sigma^\prime(k)\mathrm{d}k.
\end{split}
\end{equation*}
\hfill////

\begin{thm}
Let $\Psi$ be an absolutely continuous function
with derivative  $\Psi^{\prime}$  of polynomial growth.
Under Conditions~\ref{lpp1} and \ref{dens},
it holds that
\begin{equation*}
\mathbb{E}[\Psi(\log(S_T/F))S_T/F] = 
\int_{-\infty}^{\infty} \left\{
\Psi(g_1(z)) + \Psi^\prime(g_1(z)) - \Psi^\prime(g_2(z))e^{g_2(z)}
     \right\} \phi(z)\mathrm{d}z.
\end{equation*}
\end{thm}
{\it Proof: }
We have
\begin{equation*}
 \mathbb{E}[\Psi(\log(S_T/F))S_T/F]
= \int_{-\infty}^{\infty}
\phi(f_1(k))\left\{
\Psi(k) \frac{\mathrm{d} f_1}{\mathrm{d} k}(k)
-\Psi^\prime(k) \sigma^\prime(k)
\right\} \mathrm{d}k
\end{equation*}
by the same argument as in Theorem~\ref{mainp}.
Now, notice that
\begin{equation*}
\begin{split}
\int \Psi^\prime(g_2(z))e^{g_2(z)}\phi(z)\mathrm{d}z
=& \int
\Psi^\prime(k)e^{k}\phi(f_2(k))\frac{\mathrm{d}f_2}{\mathrm{d}k}(k)
 \mathrm{d}k\\
=& \int
\Psi^\prime(k)\phi(f_1(k))\left\{\frac{\mathrm{d}f_1}{\mathrm{d}k}(k) +
 \sigma^\prime(k) \right\}
 \mathrm{d}k\\
=& \int \Psi^\prime(g_1(z))\phi(z) \mathrm{d}z + \int \phi(f_1(k)) \Psi^\prime(k)\sigma^\prime(k)\mathrm{d}k.
\end{split}
\end{equation*}
\hfill////

\begin{rem} \upshape
The assumption that $\Psi^{\prime}$ is of polynomial growth
in the preceding theorems
can be relaxed by assuming instead 
the existence of higher moments of $S_T$ and $S_T^{-1}$ in the light of 
Theorems~3.2 and 3.4 of Lee~\cite{Lee}.
\end{rem}

\begin{rem}\upshape
The formulas in the preceding theorems 
enable us to derive directly the fair price 
of an European option from the implied volatility surface.
The point  is that no derivative of $\sigma$
is appeared in the formulas.
This is important in practice because 
the implied volatility $\sigma(k)$ is discretely observed.
The terms including the second derivative of $\Psi$
in Theorems~\ref{main} and \ref{mainp}
are approximated by  natural Riemann sums.
For the integrals with respect to $\phi(z)\mathrm{d}z$, we remark that it is not necessary to 
estimate the inverse functions $g_1$, $g_2$. For example, 
in order to use the formula in Theorem~\ref{main}, it suffices
to interpolate $(x_j,y_j)$ by a $C^1$ and piecewise $C^2$ function
$h$ so that $y_j = h(x_j)$,
where  $x_j = f_2(k_j)$, 
$y_j =  \Psi(k_j) - \Psi^\prime(k_j)(k_j+\sigma(k_j)^2/2)$.
In addition, if we take a piecewise polynomial function as  $h$,
then no numerical integration is needed due to
a well-known property of the Hermite polynomial system.
The model-free bounds on $\sigma_1, \sigma_2, g_1$ and $g_2$ given in the previous
 section should be taken into consideration in interpolating and
 extrapolating those functions.
\end{rem}

\end{document}